\documentstyle[aps,pre,graphicx]{revtex}

\begin{document}
\draft
\wideabs{
\title{Slow logarithmic relaxation in models with hierarchically constrained
 dynamics}
\author{J. J. Brey and A. Prados}
\address{F\'{\i}sica Te\'{o}rica, Universidad de Sevilla, Apartado de
  Correos 1065, 41080 Sevilla, Spain}
\date{today}
\maketitle
\begin{abstract}
A general kind of models with hierarchically constrained dynamics is shown to
exhibit logarithmic anomalous relaxation, similarly to a variety of complex
strongly interacting materials. The logarithmic behavior describes most of the
decay of the response function.
\end{abstract}
}

In the last years, an anomalous slow relaxation that can be accurately described
by a logarithmic law, has been found in the time evolution of a variety of
complex strongly interacting materials. This includes spin-glasses
\cite{NSLyS86,SyG89}, granular materials \cite{JLyN89,BSTGyK92,TSBKTyT95,MNyD92},
structural glasses \cite{TBLyK98,FGSTyT99,KyG93}, and protein models
\cite{FyA97,SGyZ98}. In all the above cases, the decay of some perturbation
$\phi(t)$ can be fitted at intermediate times by an expression of the form
\begin{equation}
\label{1.1}
\phi(t)=C_{1}-C_{2} \ln t,
\end{equation}
where $C_{1}$ and $C_{2}$ are constant, and may depend on the macroscopic
variables (temperature, compactivity, \ldots)  characterizing the state of the
system. The time range in which the logarithmic law accurately describes the
relaxation covers most of the change of $\phi(t)$ from its initial value to the
final one.

Palmer, Stein, Abrahams, and Anderson (PSAA) in a seminal paper \cite{PSAyA84}
described some dynamical models for relaxation in complex systems with strong
interactions. The models were based in a {\em hierarchically constrained
dynamics} so that the relaxation of the system involves a sequential series of
correlated processes. Although PSAA concentrate in those physical systems
showing ``stretched exponential'' relaxation, it is clear that the general
picture emerging from hierarchically constrained dynamics can be translated into
very different physical situations. As a consequence, the above models provide a
very useful tool to investigate many kinds of ``anomalous relaxation'' in
complex systems.

In particular, the picture of hierarchical constrained dynamics seem to be
physically appropriate to describe many of the complex systems showing
logarithmic relaxation. Let us consider as a particular example the
densification of powders and structural glasses at high hydrostatic pressure
studied in Refs. \cite{BSTGyK92,TSBKTyT95,TBLyK98,KyG93}. External pressure
causes some selected mesoscopic regions of the system to reach a more compact
state very quickly. The nearest neighbor regions are influenced by this
rearrangement, and can also reorganize themselves. Of course, these new
structural changes will affect the regions outside the nearest neighbors,
allowing them to relax, and so on. The above discussion suggests that the
relaxation in this kind of systems is slow because it consists of a large number
of correlated reorganization steps with an increasing characteristic time. In
other words, it has a hierarchical structure, with the faster degrees of freedom
(associated to the ``inner'' regions) controlling the relaxation of the slower
degrees of freedom (associated to the ``outer'' regions). A similar picture has
been already suggested in Refs.\ \cite{BSTGyK92,TSBKTyT95}. The aim of this
paper is to introduce a quite general class of hierarchical models, which seem
appropriate to describe this kind of systems, in order to derive the logarithmic
relaxation of Eq. (\ref{1.1}).

Like PSAA we will formulate our model in the most general way, hoping it
characterizes the behavior of a specific class of systems. We assume that the
state of the system can be given by means of $N$ modes. They do not correspond
to a particle description of the system, but to some mesoscopic level of
description. In this sense, $N$ does not scale with the size of the system, but
it is assumed to have a definite value, independent of the size, for a large
enough (macroscopic) system. Of course, the specific value depends on the system
and the physical situation we are dealing with.

The $N$ modes can be grouped into a discrete series of levels, $n=0,1,2,
\cdots$. The number of modes belonging to level $n$ will be denoted by $N_{n}$,
so that $N=\sum_{n}N_{n}$. Moreover, modes in level $n$ will be indicated by the
variables $S_{i}^{(n)}$, $i=1,2,\cdots,N_{n}$. For the densification experiments
of Refs. \cite{BSTGyK92,TSBKTyT95,TBLyK98,KyG93}, we can regard $S_i^{(n)}$ as a
pseudospin variable, with its two possible values corresponding to different
volumes of the mesoscopic regions. The dynamics of the system is defined in a
hierarchical way as follows. A mode in level $n+1$ can only modify its state if
the configuration of modes in level $n$ belongs to a well defined subset of all
the possible configurations of the modes in that level. Clearly, this slows down
the modes in a given level, as compared with those of the previous level. More
specifically, the average transition rate $W_{n+1}$ for a mode in level $n+1$ is
assumed to have the form \cite{PSAyA84}
\begin{equation}
\label{1.2}
W_{n+1}=W_{n} P_n \, ,
\end{equation}
where $P_n$ is the probability that the modes in level $n$ are in any of the
configurations allowing to relax level $n+1$. Our choice for $P_n$ will be
\begin{equation}
\label{1.2b}
P_n=\exp (-N_{n} \Delta \mu),
\end{equation}
with $\Delta\mu$ being a dimensionless free energy defining the activation
barrier per mode. Eq. (\ref{1.2b}) appears quite naturally if $P_n$ is evaluated
from the equilibrium distribution of the system under consideration, provided
that the subset of configurations of level $n$ allowing to relax level $n+1$
involve all the modes (or an extensive part of them) in level $n$. Of course,
the free energy $\Delta\mu$ must be appropriately interpreted in each particular
situation. For instance, in the densification experiments at high pressure
\cite{BSTGyK92,TSBKTyT95,TBLyK98,KyG93} it would be an activation enthalpy
divided by the temperature of the sample. On the other hand, in a granular
system, it could be understood as an ``effective volume'' divided by the
compactivity \cite{EyO89,MyE89,ByP00,PyB00}. Thus, the corresponding average
relaxation time of the mode is
\begin{equation}
\label{1.3}
\tau_{n+1}=\tau_{n} \exp (N_{n} \Delta \mu).
\end{equation}
For the sake of simplicity, the barrier $\Delta \mu$ is assumed to be
independent of the level $n$ of the mode and also of the number of modes in the
level $N_{n}$. Iteration of Eq.\ (\ref{1.3}) yields
\begin{equation}
\label{1.4}
\tau_{n}=\tau_{0} \exp \left( \Delta \mu \sum_{m=0}^{n-1}N_{m} \right).
\end{equation}

The range of variation of $n$ and the values of the populations $N_{n}$
characterize the complexity of the time evolution of the system. In principle,
there is no need for a maximum value $n_{\text{max}}$ of $n$, i.e.
$n_{\text{max}}$ may be infinite. On the other hand, the systems we are
interested in  show slow relaxation, which requires a large relaxation time. In
this way, it is necessary that
\begin{equation}
\label{1.7}
\tau_{\text{max}}=\lim_{n \rightarrow n_{\text{max}}} \tau_{n} \gg \tau_{0} \, .
\end{equation}
This condition for slow relaxation was already  pointed out by PSAA
\cite{PSAyA84}.

We are interested in the relaxation of the quantity
\begin{equation}
\label{1.5}
Q(t)=\frac{1}{N} \sum_{n} \sum_{i=1}^{N_{n}} \langle S_{i}^{(n)}(t) \rangle \, ,
\end{equation}
where the angular brackets denote configuration average. Note that in the
context of the densification experiments \cite{BSTGyK92,TSBKTyT95,TBLyK98,KyG93}
$Q(t)$ would be proportional to the actual volume of the sample at time $t$,
with the physical meaning given above to the variables $S_i^{(n)}$. The
corresponding relaxation function is defined in terms of the asymptotic long
time value $Q(\infty)$ as
\begin{equation}
\label{1.6}
\phi (t)=\frac{Q(t)-Q(\infty)}{Q(0)-Q(\infty)}=\sum_{n} \omega_{n}
\exp \left( -\frac{t}{\tau_{n}} \right),
\end{equation}
with $\omega_{n}=N_{n}/N$ being the fraction of modes in level $n$. Combination
of Eqs. (\ref{1.4}) and (\ref{1.6}) gives
\begin{equation}
\label{1.8}
\phi(t)=\sum_{n=0}^{n=n_{\text{max}}} \omega_{n} \exp \left[ -\frac{t}{\tau_{0}}
\exp\left(-\zeta\sum_{m=0}^{n-1}\omega_{m}\right) \right],
\end{equation}
where
\begin{equation}
\label{1.8b}
 \zeta=N \Delta \mu \, .
\end{equation}
Next we will assume that $\omega_{n}$ changes very smoothly with $n$. This
allows to introduce a continuum limit for the distribution of relaxation times,
which is expected to be closer to real systems than the discrete level model
considered up to now. We define a function $\tilde{\omega}(x)$ by
\( \tilde{\omega}(n\epsilon)=\omega_{n} \),
with $\epsilon \ll 1$, so that the variable $x_{n}=n\epsilon$ is almost
continuum and the sums over $n$ can be replaced by integrals. Let us introduce
the new variable ($x_{\text{max}}=n_{\text{max}}\epsilon$)
\begin{equation}
\label{1.10}
u=\frac{\int_{0}^{x}d x^{\prime} \tilde{\omega}(x^\prime)}{
  \int_{0}^{x_{\text{max}}}d x^{\prime} \tilde{\omega}(x^\prime)},
\end{equation}
$0\leq u \leq 1$, which measures the fraction of the total number of modes $N$
which are in levels up to a value $n=x/\epsilon$. It is then a simple task to
transform Eq.\ (\ref{1.8}) into
\begin{equation}
\label{1.11}
\phi(t)=\int_{0}^{1} d u\, \exp \left( -\frac{t}{\tau_{0}} e^{-\zeta u}
\right) \, ,
\end{equation}
showing that the decay of $\phi$ can be characterized by only two parameters,
namely $\tau_{0}$ and $\zeta$. The former fixes the time scale, while the latter
determines the actual shape of the relaxation. The elementary relaxation times
given by Eq.\ (\ref{1.4}) are
\( \tau(u)=\tau_{0} \exp (\zeta u) \),
and, in particular, $\tau_{\text{max}}=\tau_{0} \exp \zeta$. Thus, the condition
for slow relaxation, Eq.\ (\ref{1.7}), implies that $\exp\zeta \gg 1$.

The analysis of Eq.\ (\ref{1.11}) is greatly simplified by realizing that it is
equivalent to
\begin{equation}
\label{1.13}
\phi(t)=\frac{1}{\zeta} \left[ E_{1} \left( \frac{t}{\tau_{0}} e^{-\zeta}
\right)-E_{1} \left( \frac{t}{\tau_{0}} \right) \right],
\end{equation}
where $E_{1}(z)$ is the integral exponential function \cite{AyS65}.  There are
two natural time scales in Eq.\ (\ref{1.13}). One is that defined by
$t/\tau_{0}$, while the other is a slow time scale defined by
\( s=\frac{t}{\tau_{0}} e^{-\zeta} \ll \frac{t}{\tau_{0}} \).
In the following, we will use the asymptotic expansions of the exponential
integral function from Ref.\ \cite{AyS65}. For very short times such that
$t/\tau_{0}\ll 1$, Eq.\ (\ref{1.13}) is equal to
\begin{equation}
\label{1.15}
\phi(t) = 1-\frac{1-e^{-\zeta}}{\zeta}\frac{t}{t_{0}}
+ {\cal O}\left(\frac{t}{\tau_{0}} \right)^{2} \, ,
\end{equation}
i.e., $\phi(t)$ decays very little from its initial unity value in this region.
In the opposite limit of very large times for which $s\gg 1$, it is
\begin{equation}
\label{1.16}
\phi(t) \sim \frac{1}{\zeta} \frac{e^{-s}}{s} \, ,
\end{equation}
the response function has already decayed to very small values. One main
conclusion of the above discussion is that the relevant part of the relaxation,
in which $\phi(t)$ changes from values very close to unity to very close to
zero, takes place between the two asymptotic time regimes we have analyzed. In
this way, we are led to to consider the distinguished limit
\( t/\tau_{0} \gg 1\), \( s\ll 1 \).
In this intermediate time window it is found that
\begin{equation}
\label{1.18}
\phi(t) \sim 1-\frac{1}{\zeta} \left( \gamma+\ln \frac{t}{\tau_{0}} \right).
\end{equation}
\begin{figure}[!t]
\centerline{\includegraphics[scale=0.5]{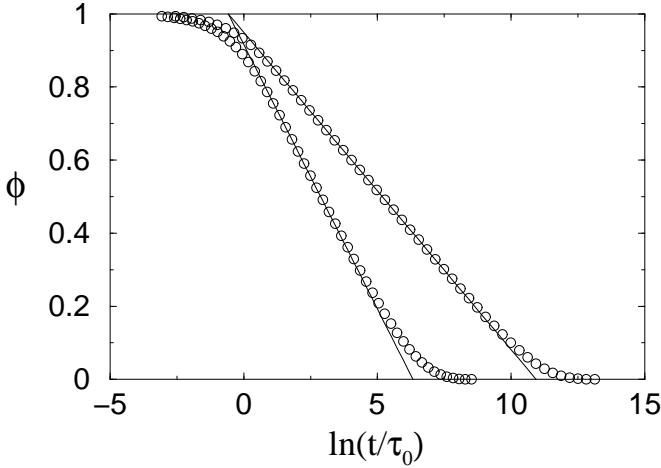}}
\caption{
Relaxation function $\phi$ plotted as a function of time. From left to right,
the curves correspond to $\exp\zeta=10^3$ and $10^5$. The integral
representation of $\phi$, Eq. (\protect\ref{1.13}), is given by the circles,
while the solid lines are the asymptotic expression for intermediate times, Eq.
(\protect\ref{1.18}).}
\label{fig1}
\end{figure}
This is precisely the logarithmic decay of Eq.\ (\ref{1.1}). Moreover, the
previous discussion of the short and large time limits, clearly indicates that
most of relaxation of the system is expected to be accurately described by Eq.\
(\ref{1.18}). As an example, we present in Fig. \ref{fig1} the relaxation
function $\phi(t)$ for $\exp\zeta=10^3$ and $\exp\zeta=10^5$. The integral
expression given by Eq.\ (\ref{1.11}) (circles) is well fitted by the
logarithmic asymptotic form of Eq.\ (\ref{1.18}) (solid lines). The agreement
extends over several decades of intermediates times, measured in units of the
basic time $\tau_{0}$. The fit improves as the value of $\zeta$ increases,
consistently with the condition for time scales separation.

It is also interesting to compute an average relaxation time $\bar{\tau}$, given
by
\begin{equation}
\label{1.19}
\bar{\tau}=\int_{0}^{\infty} dt\, \phi (t)=
\tau_{0} \frac{e^{\zeta}-1}{\zeta} \sim
\tau_{0} \frac{e^{\zeta}}{\zeta},
\end{equation}
showing an exponential increase with $\zeta$. Taking into account the definition
of $\zeta$, Eq.\ (\ref{1.8b}), the above result can be understood as
corresponding to an activation energy barrier whose height is proportional to
the complexity of the system, measured by the total number of modes $N$ actively
involved in the relaxation. As already mentioned, for a given specific
situation, the dimensionless parameter $\Delta\mu$ might have a well defined
physical interpretation, being for instance the ratio between the Gibbs free
energy barrier per mode and the thermal energy. In this way, it is also
interesting to note that the coefficient of $\ln t$ in Eq.\ (\ref{1.18}) is
$\zeta^{-1}$, which is then proportional to the temperature $T$. This would lead
to the $T \ln t$ behavior observed in many physical systems, for instance in
Refs.\ \cite{SyG89,KyG93}.

In summary, we have studied a simple class of models which exhibit an anomalous
logarithmic relaxation. The dynamics is hierarchically constrained, and they are
adequate to study slow relaxation because there is a clear separation  of time
scales. The slow logarithmic relaxation is similar to what is experimentally
observed in a variety of complex systems, fitting the data in most of the
relevant part of the relaxation, i.e., the time window where the relaxation
function is neither too small nor too close to its initial value. Perhaps the
main assumption of this work is the ``extensiveness'' with $N_n$ of the
effective barrier for modes in level $n+1$, as expressed by Eq.\ (\ref{1.2b}).
This is quite a general condition and, therefore, the class of hierarchical
models presented here may provide an explanation of the appearance of linear
logarithmic relaxation in very different physical situations. In this context,
the role of hierarchically constrained dynamics may also be relevant in order to
understand other kinds of anomalous relaxation, as the inverse logarithmic
behavior found in compaction experiments of granular systems
\cite{KFLJyN95,NKBJyN98}. Also, it seems interesting to investigate the behavior
of hierarchical models in the temperature cycling experiments usually carried
out in glassy systems \cite{ByPunpub}.

\acknowledgments
This research was partially supported by Grant No. PB98-1124 from the
Direcci\'{o}n General de Investigaci\'{o}n Cient\'{\i}fica y T\'{e}cnica
(Spain).

\end{document}